\documentclass[aps,twocolumn,showpacs,superscriptaddress,groupedaddress]{revtex4-1}  
\usepackage{graphicx}  
\usepackage{dcolumn}   
\usepackage{bm}        
\usepackage{amssymb}   

\begin{document}



\title{Effect of Coriolis force on homoclinic bifurcations in thermal convection}
\date{\today}
\author{Priyanka Maity }
\author{Krishna Kumar}
\affiliation{Department of Physics and Meteorology, IIT Kharagpur, Kharagpur- 721302, India}
\author{Pinaki Pal}
\affiliation{Department of Mathematics, National Institute of Technology, Durgapur-713 209, India}
\begin{abstract}
We present the effects of small Coriolis force on homoclinic gluing and ungluing bifurcations in low-Prandtl-number($0.025 \leq Pr$) rotating Rayleigh-B\'enard system with \textit{stress-free} top and bottom boundaries. We have performed direct numerical simulations for a a wide range of Taylor number ($5 \leq Ta \leq 50$) and reduce Rayleigh number $r$ ($ \leq 1.25$). We observe homoclinic ungluing bifurcation, marked by the spontaneous breaking of a larger limit cycle into two possible set of limit cycles in the phase space, for lower values of $Ta$. Two unglued limit cycles merge together for higher values of $Ta$, as $Ra$ is raised sufficiently. The range of $Ta$ for which both gluing and ungluing bifurcation can be seen depends on the Prandtl number $Pr$. The variation of the bifurcation points with $Ta$ is also investigated. We also present a low-dimensional model which qualitatively captures the dynamics of the system near the homoclinic bifurcation points. The model is used to study the variation of the homoclinic bifurcation points with Prandtl number for different values of $Ta$.
\end{abstract}

\pacs{47.20.Ky, 47.27.ek, 47.57.-r}
\maketitle

\section{Introduction}
Nonlinear extended dissipative systems, driven away from equilibrium,  reveal a wide range of bifurcations and pattern dynamics~\cite{pattern_formation}. A homoclinic gluing~\cite{gluing} occurs when two limit cycles simultaneously become homoclinic orbits to a single saddle point. This leads to spontaneous merging of two limit cycles to a single limit cycle in the appropriate phase space, as the bifurcation parameter is raised above a critical value. The gluing bifurcation is observed in liquid crystals~\cite{gluing_liquid_crystals}, in fluids~\cite{gluing_fluids,pal_etal_2013}, and in electronics circuits~\cite{gluing_circuits}. Recently, Pal~\textit{et. al}~\cite{pal_etal_2013} showed the possibility of a homoclinic bifurcation in very low-Prandtl-number thermal convection and investigated the pattern dynamics near the bifurcation point. 

Rayleigh-B\'enard convection (RBC), where a horizontal layer of a fluid is subjected to an adverse temperature gradient, has been widely studied as a model system to investigate interesting pattern dynamics~\cite{patterns_RBC}, bifurcations~\cite{bifurcation}, route to chaos~\cite{route_to_chaos} and turbulence~\cite{turbulence}. Low-Prandtl-number convection~\cite{lowP, zeroP, kft}, which is relevant in the geophysical~\cite{geophys} and the astrophysical~\cite{astrophys} context, show interesting pattern dynamics~\cite{lowp_patterns} and a plethora of bifurcations. A uniform rotation about a vertical axis~\cite{chandrasekhar_book, rossby_1969}, on the other hand,  introduces a Coriolis force in the system that breaks the mirror symmetry of the patterns and show several patterns~\cite{rotating_patterns}. However, the effect of Coriolis force on homoclinic bifurcations has not been explored. 

We present in this article the effect of slow rotation on homoclinic bifurcations and pattern dynamics in its vicinity in low-Prandtl-number RBC using direct numerical simulations (DNS) and a low dimensional model. We observed spontaneous breaking (ungluing) of a larger limit cycle into two limit cycles near a homoclinic point for lower values of Taylor number $Ta$ and again spontaneous merger (gluing) of two limit cycles into one at higher values of $Ta$. We also present a simple model and use it to investigate the Prandtl number dependence of the homoclinic point.

\begin{table*}
  \begin{center}
\def~{\hphantom{0}}
  \begin{tabular}{c|c|c|c|c}
\hline
 Convective  &\multicolumn{2}{c|}{$Pr = 0$}&\multicolumn{2}{c}{$ Pr = 0.025$} \\
\cline {2-3} \cline{4-5}
patterns  &r(Ta = 10)&  r(Ta = 40) & r(Ta = 10) & r(Ta = 25)\\ 
\hline\hline
STO	&1.001 - 1.018~& 1.001  -1.071~& 1.001 -  1.02~& 1.001 - 1.021 \\
\hline
STC	&1.019 - 1.088~& 1.072 - 1.116~& 	    -   	 ~& 1.022 - 1.037\\
\hline
OCR -II$^{\prime}$& - & 1.117 - 1.127 & - & 1.038 - 1.139 \\
\hline
OCR-I	&1.089 - 1.120~& 1.128 - 1.154 ~& 1.051 - 1.148~& 1.140 - 1.175\\
\hline
OCR-II	&1.121 - 1.137~& 1.155 - 1.163~& 1.149 -  1.168~& 1.176 - 1.186\\
\hline
CR	&1.138 - 1.187~& 1.164 - 1.183~& 1.170 -  1.210~& 1.187 - 1.209\\
\hline
SQ	& $\leq 1.189$~& $\leq 1.184$~& $\leq 1.211$ & $\leq 1.210$ \\
\hline
 \end{tabular}
\caption{Flow patterns observed in DNS for different regions of $r$ for two values of $Pr$ . The patterns are: Self-tuned periodic competition of rolls (STO), chaotic competition of self-tuned rolls (STC), oscillating cross-rolls  with $|W_{101}| = |W_{011}|$ (OCR-I), oscillating cross-rolls  with $|W_{101}|_{max} \neq |W_{011}|_{max}$ (OCR-II and OCR -II$^{\prime}$), stationary cross-rolls (CR) ($W_{101} = W_{011}$) and stationary squares (SQ) ($W_{101} \neq W_{011}$).}
\label{tab:flow_pattern}
 \end{center}
 \end{table*}

\section{The hydrodynamic system}
We consider a thin horizontal layer of Boussinesq fluid of thickness $d$, kinematic viscosity $\nu$, thermal expansion coefficient $\alpha$, thermal diffusivity $\kappa$, rotating slowly with a uniform angular velocity {\boldmath $\Omega$} about a vertical axis, and subjected to an adverse temperature gradient $\beta$ in the vertical direction. The hydrodynamics of the rotating Rayleigh-B\'{e}nard convection is governed by,
 
\begin{eqnarray}
\partial_{t}\boldsymbol{v}+(\boldsymbol{v\cdot\nabla})\boldsymbol{v} &=& 
    -\boldsymbol{\nabla}p + Ra\theta{\boldsymbol \lambda} \nonumber \\
    &+& \nabla^{2}\boldsymbol{v}+\sqrt{Ta} ({\boldsymbol{v\times \lambda}}),\label{ns}\\
Pr[\partial_{t}\theta+(\bf{v\cdot\nabla})\theta] &=& \nabla^{2}\theta + v_{3},  \label{temp}\\
 \boldsymbol{\nabla \cdot v} &=& 0,  \label{cont}
\end{eqnarray}
\noindent where $\boldsymbol{v}(x,y,z,t)$ $\equiv$ $(v_{1},v_{2},v_{3})$, $\theta(x,y,z,t)$ and $p(x,y,z,t)$ are   respectively the flow velocity, the convective temperature field and the convective pressure field. The unit vector ${\boldsymbol \lambda}$ is directed opposite to the direction of the gravitational acceleration ${\boldsymbol{g}}$. In the above equations all the length scales are made dimensionless by the fluid thickness $d$, the time scale by the viscous diffusive time $d^2/\nu$, and the temperature field by $\nu \beta d /\kappa$. The dimensionless parameters are: Rayleigh number $Ra = (g \alpha \beta d^4)/(\nu \kappa)$, Prandtl number $Pr = \nu/\kappa$ and Taylor number $Ta = 4 \Omega^2 d^4/\nu^2$. In the limit $Pr \to 0$, eq.~\ref{temp} is slaved to the vertical velocity, i.e., $\nabla^{2}\theta = - v_{3}$.
We consider thermally conducting and \textit{stress-free} bounding surfaces. This leads to the boundary conditions $\theta$ $=$  $\partial_{z}v_{1}$ $=$ $\partial_{z}v_{2}$ $=$ $v_{3} = 0$ on the boundaries located at $z=0$ and $1$. All the convective fields are assumed to be periodic in horizontal plane. The direct numerical simulations (DNS) of the hydrodynamic system (eqs.~\ref{ns}-\ref{cont}) with the boundary conditions is carried out using an open-source code TARANG~\cite{tarang_2011}  based on pseudo spectral method. All the convective fields are expanded as:
\begin{eqnarray}
{\bf \Psi} (x,y,z,t) &=& \sum_{l,m,n} \Psi_{lmn}(t) e^{ik(lx+my)} f_n(z),\label{eq.DNS_expansion},\nonumber\\
{\bf \Phi} (x,y,z,t) &=& \sum_{l,m,n} \Phi_{lmn}(t) e^{ik(lx+my)} g_n(z),\label{eq.DNS_expansion},
\end{eqnarray}
\noindent where ${\bf \Psi} (x, y, z, t) \equiv (v_1, v_2, p)^T$, $f(z) = \sin{(n \pi z)}$ and ${\bf \Phi} (x, y, z, t) \equiv (v_3, \theta)^T$, $g(z) = \cos{(n \pi z)}$. We consider a periodic box of size $2\pi/k: 2\pi/k : 1$ with $k = k_{c} (Ta)$ for our simulations.  $k_c (Ta)$ is the critical wave number predicted by linear theory~\cite{chandrasekhar_book}. The spatial grid resolution of $64 \times 64 \times 64$ is used for DNS, which is quite good to resolve the flow structure near onset. The time advancement is done using standard fourth order Runge-Kutta (RK4) integration scheme. The time step used for integration varies in the range $0.0005 \leq dt \leq 0.001$. We started our simulations with random initial conditions by fixing the values of Prandtl number and Taylor number. The reduced Rayleigh number $r = Ra/Ra_c(Ta)$, where $Ra_c$ is the critical Rayleigh number predicted by the linear theory of Chandrasekhar~\cite{chandrasekhar_book}, is increased from 1 in small steps of $0.001 \leq \Delta r \leq 0.01$. Reduced Rayleigh number has been varied  in the range $1 \leq r \leq 1.25$ for all values of Ta mentioned above. The values at the final time step has also been used as the initial condition for the next run. We have performed our runs for Taylor numbers $2 \leq Ta \leq 50$ for various values of Prandtl numbers. The case of Pr$ \to 0$ and Pr = 0.025 have been studied in details. We have also performed runs by starting with a higher value of r and then decreasing the r in small steps. We have not observed any hysteresis in the range of r considered here.

\begin{figure}[h]
\includegraphics[height=6 cm,width=8 cm]{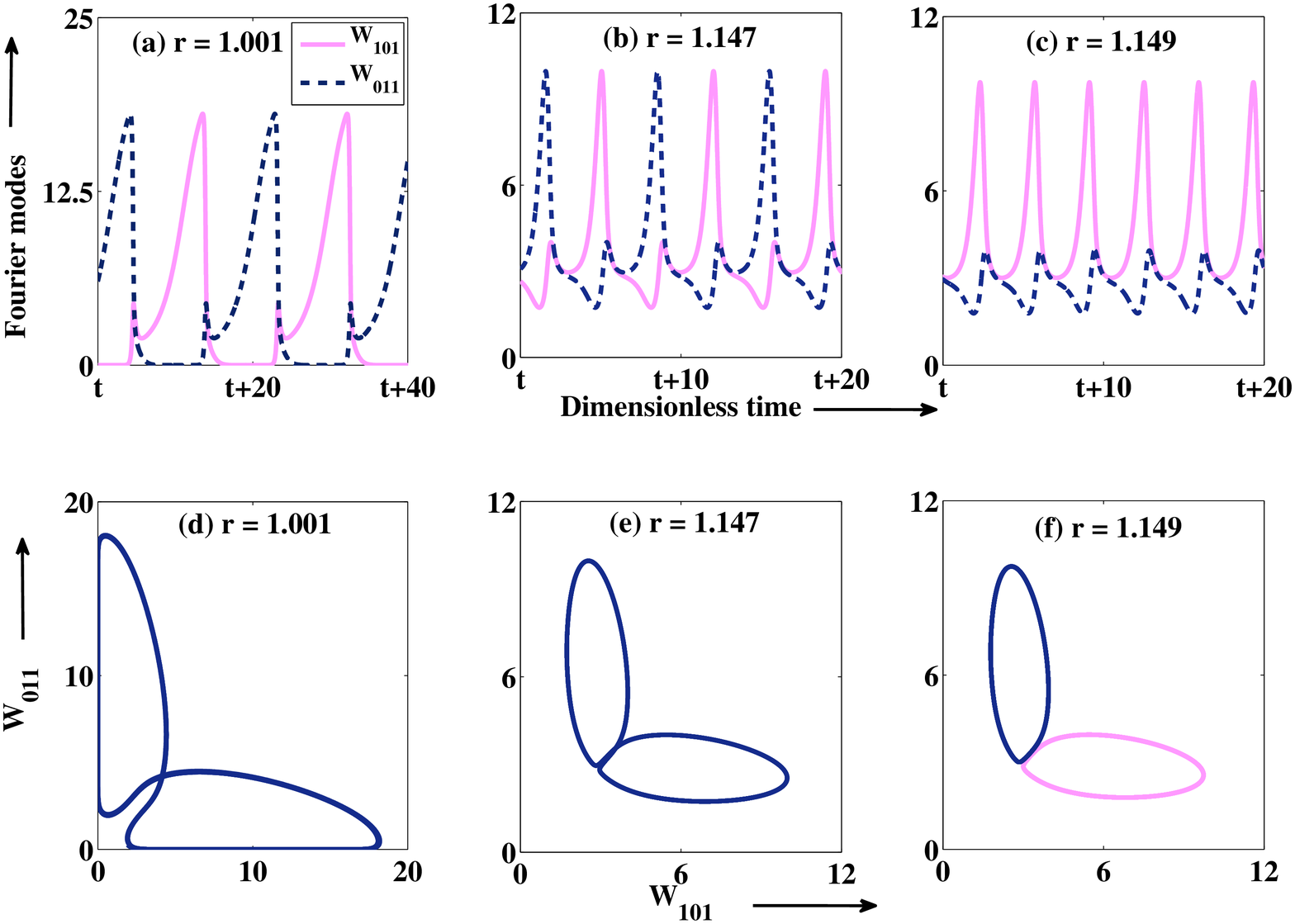}
\caption{(Colour online) Behaviour near the homoclinic bifurcation for Ta = 10 at Pr = 0.025 from DNS: The temporal variation of Fourier modes $W_{101}$ (pink (gray) solid lines) and $W_{011}$ (blue (black) dotted lines) for (a) r = 1.001, (b) r = 1.147, and (c) r = 1.149. The onset is non-local solution with a periodic competition between two mutually perpendicular set of self-tuned rolls(STO). The corresponding phase portraits on $W_{101}-W_{011}$ plane are plotted in (d), (e), and (f) respectively.}
\label{fig:DNS_T10}
\end{figure}

The various flow patterns observed for  two different values of Prandtl numbers (Pr = 0 and 0.025) for different values of Ta are listed in table~\ref{tab:flow_pattern}. The onset of convection is always self-tuned oscillations (STO) of two mutually perpendicular set of rolls. The temporal variation of the leading Fourier modes ($W_{101}$ and $W_{011}$) in STO, OCR-I, and OCR-II states at Ta = 10 and Pr = 0.025 are plotted in first row of fig.~\ref{fig:DNS_T10}. In STO state both the modes ($W_{101}$ and $W_{011}$) oscillate with zero minima and a phase difference (fig.~\ref{fig:DNS_T10} (a)). The mode $W_{101}$ (rolls along y-axis) remains zero as the mode $W_{011}$ (rolls along y-axis) grows from a minimum value. When $W_{011}$ grows to a large value $W_{101}$ starts growing and simultaneously $W_{011}$ drops to a minimum. $W_{011}$ remains zero within a period when $W_{101}$ is growing and again starts growing from the minima in the vicinity of the time when $W_{101}$ attains a maxima. This self tuning repeats periodically and is similar to the self tuning of rolls observed by Kumar, Fauve, and Thual~\cite{kft}. As r is raised, periodic self tuned oscillations become chaotic (STC) in nature. With further increase in r we reach the oscillating cross rolls (OCR-I, $|W_{101}| = |W_{011}|$) state where two mutually perpendicular set of cross-rolls start oscillating with non-zero minimum (see fig.~\cite{fig:DNS_T10} (b)). Both the STO and OCR-I states represent nonlocal solutions, with existence of one large limit cycle (as shown in figs.~\cite{fig:DNS_T10} (d) and (e)). When r is increased further beyond a certain critical value $r_{ug}$ we reach OCR-II state. In this state, the mutually perpendicular set of oscillating cross rolls break into two possible set of oscillating cross roll solutions (one with $|W_{101}| > |W_{011}|$ and another with $|W_{101}| < W_{011}$). The temporal variation of the Fourier modes in this state is shown in fig.~\ref{fig:DNS_T10} (c) and the corresponding phase portrait is shown in fig.~\ref{fig:DNS_T10} (f). We observe the existence of two limit cycles (represented by the blue (black) and pink (gray) in fig.~\ref{fig:DNS_T10} (f)) in OCR-II state. The transition from a larger limit cycle to either of the two possible smaller limit cycles is via homoclinic ungluing bifurcation with a divergence in time period of oscillation near the bifurcation. The two limit cycle gradually diminishes in size with the increment of r and ultimately reaches the stationary cross roll (CR, $W_{101} \neq W_{011}$) solutions when r is raised above a critical value $r_1$. The transition OCR-II $\rightarrow$ CR is via an inverse Hopf bifurcation. The CR state further bifurcates to the stationary square (SQ, $W_{101} = W_{011}$) solution when r exceeds the critical value $r_2$ via an inverse pitchfork.

\begin{figure}[h]
\includegraphics[height=5 cm,width=8 cm]{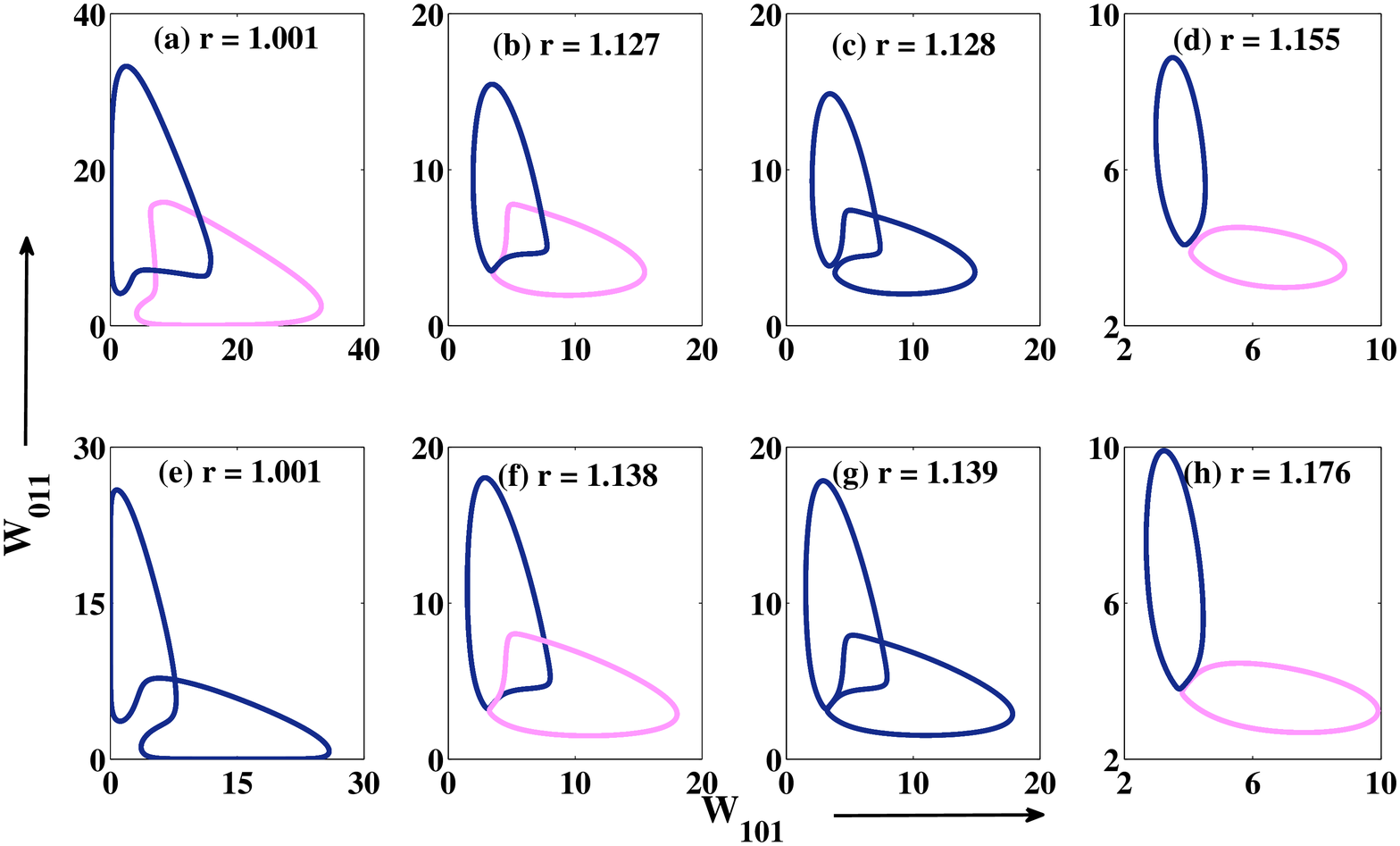}
\caption{(Colour online) Phase portraits displaying homoclinic gluing and homoclinic ungluing bifurcations  from DNS: The first row shows the phase projections for Pr = 0 at Ta = 40 for (a) r = 1.001 , (b) r = 1.127, (c) r = 1.128 and (d) r = 1.155. The second row shows the phase projections for Pr = 0.025 and Ta = 25 for (e) r = 1.001, (f) r = 1.138, (g) r = 1.139 and (h) r = 1.176.}
\label{fig:DNS_modes_ungluing}
\end{figure}

For higher values of Ta, however we observe the presence of both homoclinic ungluing and homoclinic gluing occurring with variation of r. The presence of both gluing and ungluing bifurcation is observed in the range $40 \leq Ta \leq 50$ for Pr = 0, and for $25 \leq Ta \leq 35$ for Pr = 0.025.  Figure~\ref{fig:DNS_modes_ungluing} describes the phase portraits on $W_{101}-W_{011}$ plane showing the occurrence of successive homoclinic gluing and ungluing bifurcation. The upper row of figure~\ref{fig:DNS_modes_ungluing} is for Pr = 0 at Ta = 40 and the lower row is for Pr = 0.025 at Ta = 25. The onset is self-tuned (STO) as observed in lower values of Ta. For Pr = 0 case the onset is local solution with presence of two limit cycles as shown in fig.~\ref{fig:DNS_modes_ungluing}(a). Whereas for Pr$\neq 0$ the onset is non-local solution with existence of one larger limit cycle (fig.~\ref{fig:DNS_modes_ungluing}(e)) as observed for lower values of Ta previously. The phase projections of STO, OCR-II$^{\prime}$, OCR-I, and OCR-II states on $W_{101}-W_{011}$ plane are displayed in first, second, third, and fourth column of fig.~\ref{fig:DNS_modes_ungluing} respectively. With increment of r the system enter into chaotic state and the next ordered state appears in the form of OCR-II$^{\prime}$ solution with further increase of r. The OCR-II$^{\prime}$ solution is similar to OCR-II state, with presence of two limit cycles marked by blue (black) and pink (gray) lines, described earlier but is not present for lower values of Taylor numbers. In this state the modes oscillates $W_{101}$ and $W_{011}$ oscillates with a non zero minimum and unequal amplitudes ($|W_{101}| \neq |W_{011}|$) with a constant phase difference. As r is increased above a critical value $r_g$, the two limit cycles spontaneously merge to form a bigger limit cycle and the system bifurcates to OCR-I state via a homoclinic gluing bifurcation. All the subsequent bifurcations are similar to he sequence observed for lower values of Ta described earlier.

\begin{figure}[h]
\includegraphics[height=4 cm,width=8 cm]{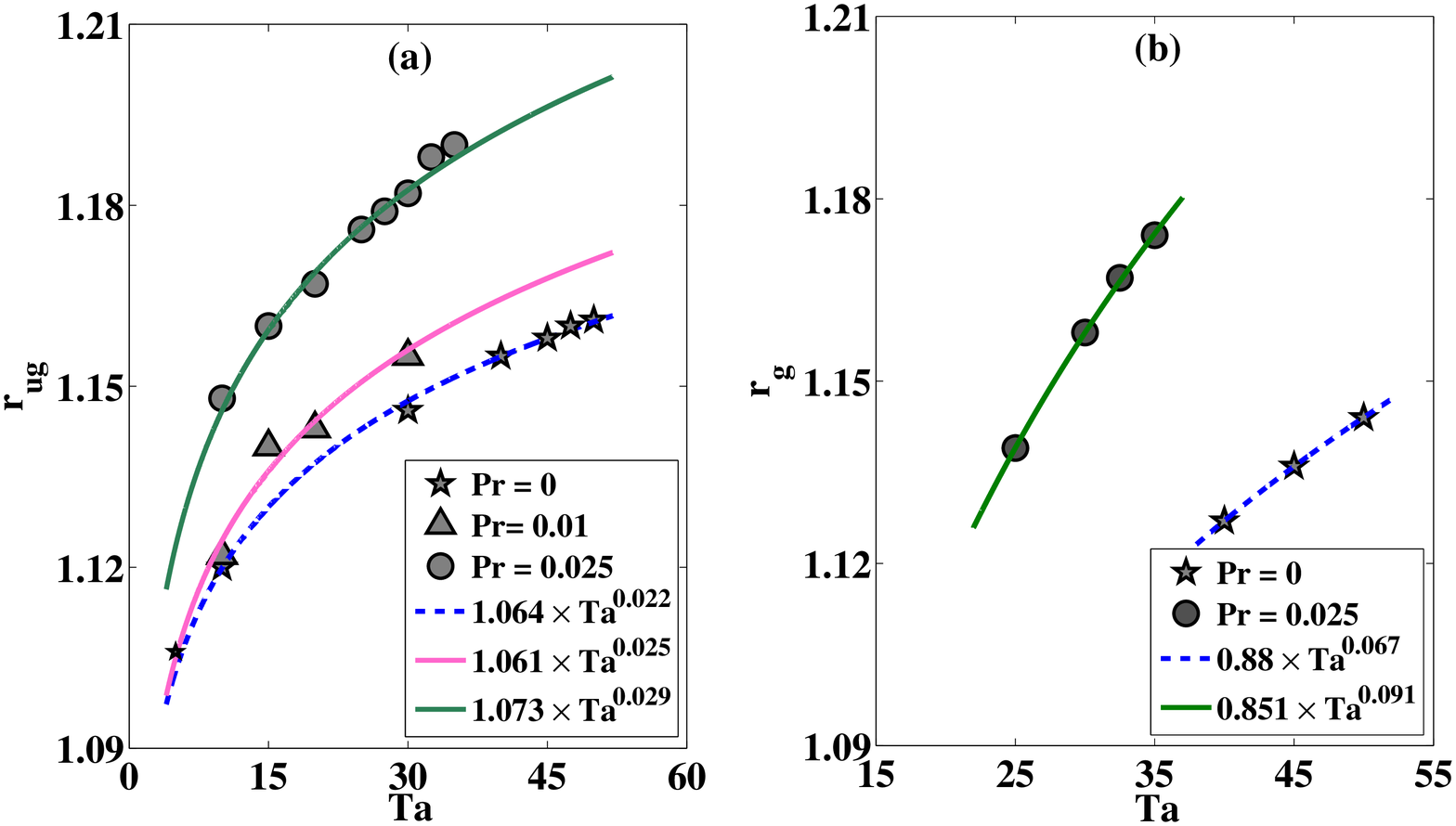}
\caption{(Colour online) Variation of homoclinic bifurcation points with Taylor number Ta as obtained from DNS: The variation of $r_{ug}$ with Ta for three values of Prandtl numbers (Pr = 0, Pr = 0.01, and Pr = 0.025) are plotted in (a). The points marked by `$\circ$' , `$\triangle$' and `$\star$' represent the data for Pr = 0, Pr = 0.01 and Pr = 0.025 respectively. Variation of homoclinic gluing bifurcation point $r_g$ with Ta for Pr = 0 and Pr = 0.025 is plotted in (b). Both $r_{ug}$ and $r_g$ shows power law dependence on Ta.}
\label{fig:DNS_scaling}
\end{figure}

The value of r at which homoclinic bifurcations occurs depends on Taylor number for a fixed Prandtl number. The dependence of homoclinic bifurcation points on Ta is plotted in fig.~\ref{fig:DNS_scaling}. The homoclinic ungluing bifurcation point $r_{ug}$, where transition from OCR-I $\rightarrow$ OCR-II take place, shows power law dependence on Ta (fig.~\ref{fig:DNS_scaling}(a)). The point $r_{ug}$ varies as Ta$^{\gamma}$ with $\gamma = 0.022$ for Pr =0,  $\gamma = 0.025$ for Pr = 0.01, and $\gamma = 0.029$ for Pr = 0.025. The transition from OCR-II$^{\prime} \rightarrow$ OCR-I occurs at $r = r_g$ only for higher values of Ta. Figure~\ref{fig:DNS_scaling}(b) shows the variation in $r_g$ as Taylor number is varied gradually. The bifurcation point $r_g$ also shows power law dependence on Ta with $r_g \propto Ta^{\beta}$. The values of $\beta$ for Pr = 0 and Pr = 0.025 are 0.067 and 0.091 respectively. The value of Taylor number upto which gluing or ungluing can be observed also depend on the value of Pr. Presence of gluing or ungluing bifurcation is observed only for $Ta \leq 50$ at Pr = 0 and for $Ta \leq 35$ at Pr = 0.025. 

\section{The model}

\begin{figure}[h]
\includegraphics[height=6 cm,width=8 cm]{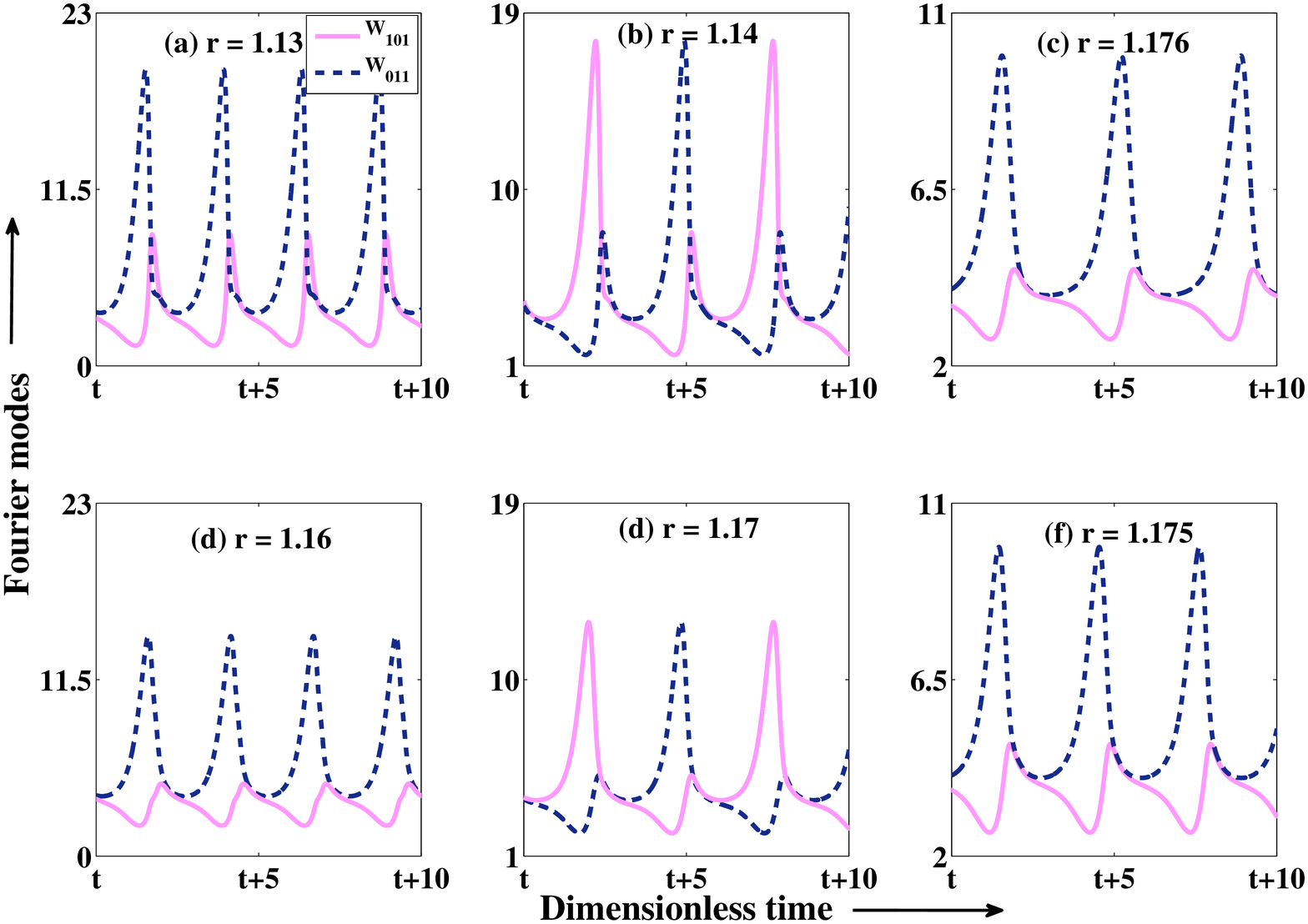}
\caption{(Colour online) Comparison of DNS and model for Pr =0.025 and Ta = 25: The temporal behaviour of leading Fourier modes $W_{101}$ (pink (gray) solid lines) and $W_{011}$ (blue (black) dotted lines) for (a) r = 1.13, (b) r = 1.14 and (c) r = 1.176 as obtained from DNS are plotted in upper row of the figure. The lower row  displays the time signal as obtained from model for (d) r = 1.16, (e) r = 1.17, and (f) r = 1.175. The model can qualitatively captures the temporal variation of the modes in OCR-II$^{\prime}$, OCR-I, and OCR-II state (represented in first, second, and third column respectively).}
\label{fig:DNS_modelcomp}
\end{figure}

Direct numerical simulations are enormously time consuming and requires huge computation facilities. Hence we try to capture the system using a few mode model, derived using Galerkin method~\cite{kft}. For this purpose we expand the convective fields as sine and cosine functions in compatible with the boundary conditions as follows:
\begin{eqnarray}
{\Phi} &=& \sum_{l,m,n} [{\Phi}_{lmn}(t)\cos{lk_cx}\cos{mk_cy} \nonumber \\
& +& {\Phi}_{\bar{l}\bar{m}n} (t)\sin{lk_cx} \sin{mk_cy}] g(z), \label{expansion_model}
\end{eqnarray}
\noindent where, $g(z) = sin(n \pi z)$ for vertical velocity $v_3$ and convective temperature field $\theta$ expansion and $g(z) = cos(n \pi z)$ for vertical vorticity ($\omega_3 = (\boldsymbol{\nabla} \times \boldsymbol{v})\cdot \boldsymbol{\lambda$}) expansion. We choose the following eight real vertical velocity  ($v_{3}$) modes: $\tilde{W}_{101}$, $\tilde{W}_{011}$, $\tilde{W}_{112}$, $\tilde{W}_{\bar{1}\bar{1}2}$, $\tilde{W}_{211}$, $\tilde{W}_{\bar{2}\bar{1}1}$, $\tilde{W}_{121}$, $\tilde{W}_{\bar{1}\bar{2}1}$, twelve real vertical vorticity (${\omega}_3$) modes : $\tilde{Z}_{101}$, $\tilde{Z}_{011}$, $\tilde{Z}_{112}$, $\tilde{Z}_{\bar{1}\bar{1}2}$, $\tilde{Z}_{211}$, $\tilde{Z}_{\bar{2}\bar{1}1}$, $\tilde{Z}_{121}$, $\tilde{Z}_{\bar{1}\bar{2}1}$,$\tilde{Z}_{\bar{1}\bar{1}0}$, $\tilde{Z}_{200}$, $\tilde{Z}_{020}$, and $\tilde{Z}_{\bar{2}\bar{2}0}$, and nine real modes for temperature field $\theta$: $\tilde{\theta}_{101}$, $\tilde{\theta}_{011}$, $\tilde{\theta}_{112}$, $\tilde{\theta}_{\bar{1}\bar{1}2}$, $\tilde{\theta}_{211}$, $\tilde{\theta}_{\bar{2}\bar{1}1}$, $\tilde{\theta}_{121}$, $\tilde{\theta}_{\bar{1}\bar{2}1}$, and $\tilde{\theta}_{002}$. We then project the expansion on eqs.~\ref{ns} and ~\ref{temp} to obtain a twenty nine mode model consisting of twenty-nine coupled first  ordered non-linear differential equations. In the limit Pr $\rightarrow$ 0, the temperature modes $\theta$ becomes slaved to v$_3$ via $\nabla^{2}\theta = - v_{3}$ and the model reduces to a twenty mode model except the $\theta$ modes.

\begin{figure}[h]
\includegraphics[height=6 cm,width=8cm]{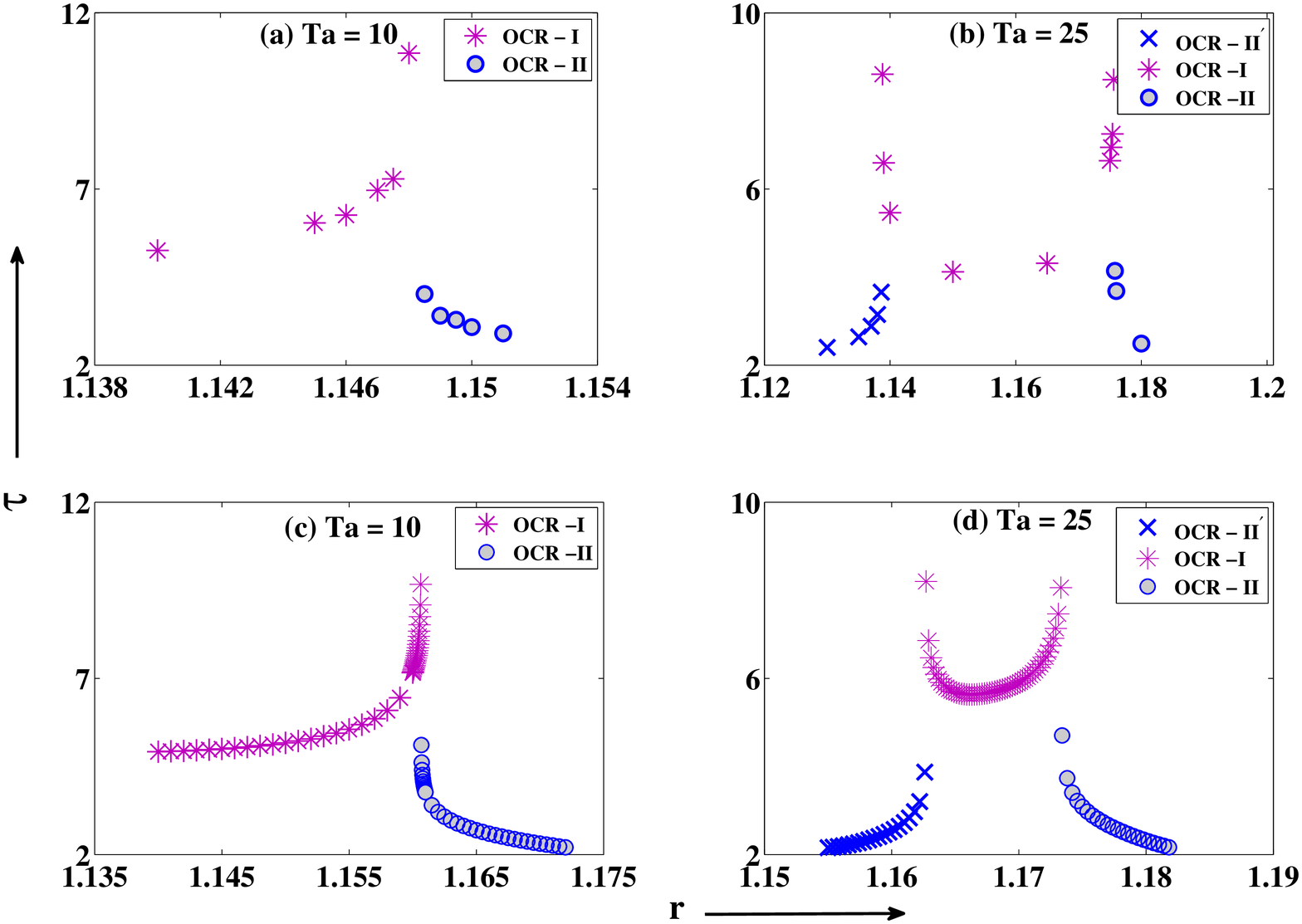}
\caption{(Colour online) Variation of dimensionless time period ($\tau$) with r near the homoclinic bifurcation points:(a) Ta = 10 and (b) Ta = 25 shows the divergence of time period of oscillations near the homoclinic bifurcation points as obtained from DNS for . The lower row of the figure shows similar behaviour for (c) Ta = 10 and (d) Ta =25 as obtained from model. The data for OCR-II$^{\prime}$, OCR-I, and OCR-II are denoted by points marked as `$\times$', `$\ast$', and `$\circ$' respectively. The data from DNS and model are in good agreement with each other.}
\label{fig:timeperiod}
\end{figure}

The model can capture all the bifurcations observed in DNS qualitatively well. The model is quite good for lower values of Taylor number and shows deviation from DNS for higher values. We observe presence ungluing bifurcation in the model for $2 \leq Ta \leq 50$ at Pr = 0 and for $2 \leq Ta \leq 25$
for Pr = 0.025. But homoclinic gluing bifurcation is observed only for higher values of Ta which is consistent with DNS. The model shows gluing bifurcation for $35 \leq Ta \leq 50$ at Pr = 0 and for $15 \leq Ta \leq 25$ for Pr = 0.025. Figure~\ref{fig:DNS_modelcomp} compares the time series obtained from DNS and the model for Ta = 25 at Pr = 0.025. The upper row of fig~\ref{fig:DNS_modelcomp} shows the temporal behaviour of the modes $W_{101}$ and $W_{011}$ in OCR-II$^{\prime}$ (fig.~\ref{fig:DNS_modelcomp}(a)), OCR-I (fig.~\ref{fig:DNS_modelcomp}(b)), and OCR-II (fig.~\ref{fig:DNS_modelcomp}(c)) as obtained from DNS. The corresponding time series from model is plotted in the lower row of the figure. The model can qualitatively capture the temporal behaviour as obtained from DNS. Figure~\ref{fig:timeperiod} displays the variation in time period of oscillations ($\tau$) with r as observed from DNS (upper row) and model (lower row) at Pr = 0.025 for Ta = 10 and Ta =25. The time period diverges near the homoclinic bifurcation points. 

\begin{figure}[h]
\includegraphics[height=6 cm,width=8 cm]{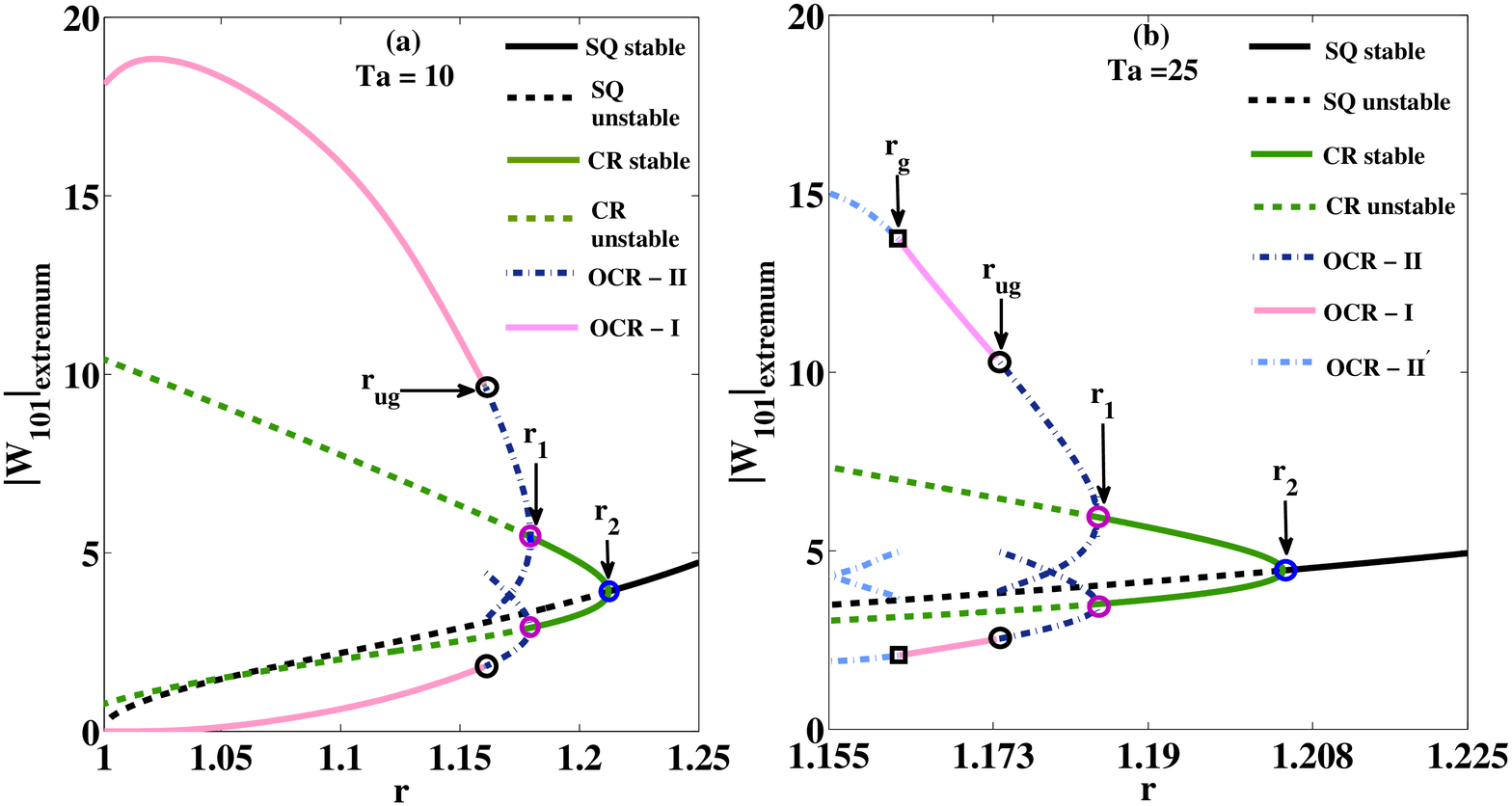}
\caption{(Colour online) Linear stability analysis for Pr = 0.025 at Ta = 10 and Ta =25: (a) displays the bifurcation diagram with variation of r as the parameter for Ta = 10. It captures STO, OCR-I, OCR-II, CR, and SQ states as r is varied in small steps. The bifurcation diagram obtained for Ta = 25 is plotted in (b). The bifurcation diagram captures OCR-II$^{\prime}$ solution in addition to all the other state mentioned in (a). The transition from OCR-II$^{\prime}$ $\rightarrow$ OCR-I takes place at $r = r_g$(represented by black square in the figure)}
\label{fig:bif_model}
\end{figure}

Now we use the model to capture the sequence of bifurcations observed in DNS. The bifurcation diagram for Ta = 10 at Pr = 0.025 is plotted in fig.~\ref{fig:bif_model}(a). The model shows existence of all the flow patterns observed in DNS except the self-tuned chaotic region (STC).  We observe two types of fixed point solutions, SQ ($|W_{101}| = |W_{011}|$) and CR ($|W_{101} \neq |W_{011}|$) in the bifurcation diagram. The SQ solutions are displayed by green (dark gray) lines, while CR solution are marked by the black lines. The stable and unstable fixed point solutions are represented by the solid and dashed lines respectively. Square fixed point are observed in the range $1.212 \leq \leq 1.25 $ and they becomes unstable for $1.212 < r$, but continues to exist as saddle point till r = 1. Stationary cross rolls solutions are stable for $1.179 \leq r \leq 1.211$ and becomes unstable for $r < 1.179$. Thus, we do not observe any fixed point solution at the onset. 

The onset of convection is self tuned oscillations with $|W_{101}| = |W_{011}|$. The STO solutions converts to OCR-I solutions as r is increased and the minimum of oscillation becomes non-zero. The global extrema of STO and OCR-I solutions are represented by the pink (light gray) solid lines. This transition takes place approximately for $1.05 \leq r$. As r is increased gradually the single limit cycle in OCR-I state comes closer and closer to the saddle fixed point. When r is raised above the critical value $r_{ug}$ = 1.161 (marked by the blue (black) circle in fig~\ref{fig:bif_model}(a)) the system bifurcates to OCR-II state. The two banches of  OCR-II solutions are represented as blue (black) dashed-dotted lines. At the the bifurcation point the limit cycle becomes homoclinic orbit to the saddle fixed point and the time period diverges. The bigger limit cycle spontaneously breaks into two smaller limit cycle and the system evolves to OCR-II sate. The  bifurcation from OCR-I to OCR-II takes via homoclinic ungluing bifurcation. Both the existing limit cycles shrink in size as r is raised higher and the system bifurcates to CR takes place at $r = r_1$ via inverse Hopf bifurcation. The CR state further bifurcates to SQ solutions at $r = r_2$ via an inverse pitchfork.

\begin{figure}[h]
\includegraphics[height=5 cm,width=8 cm]{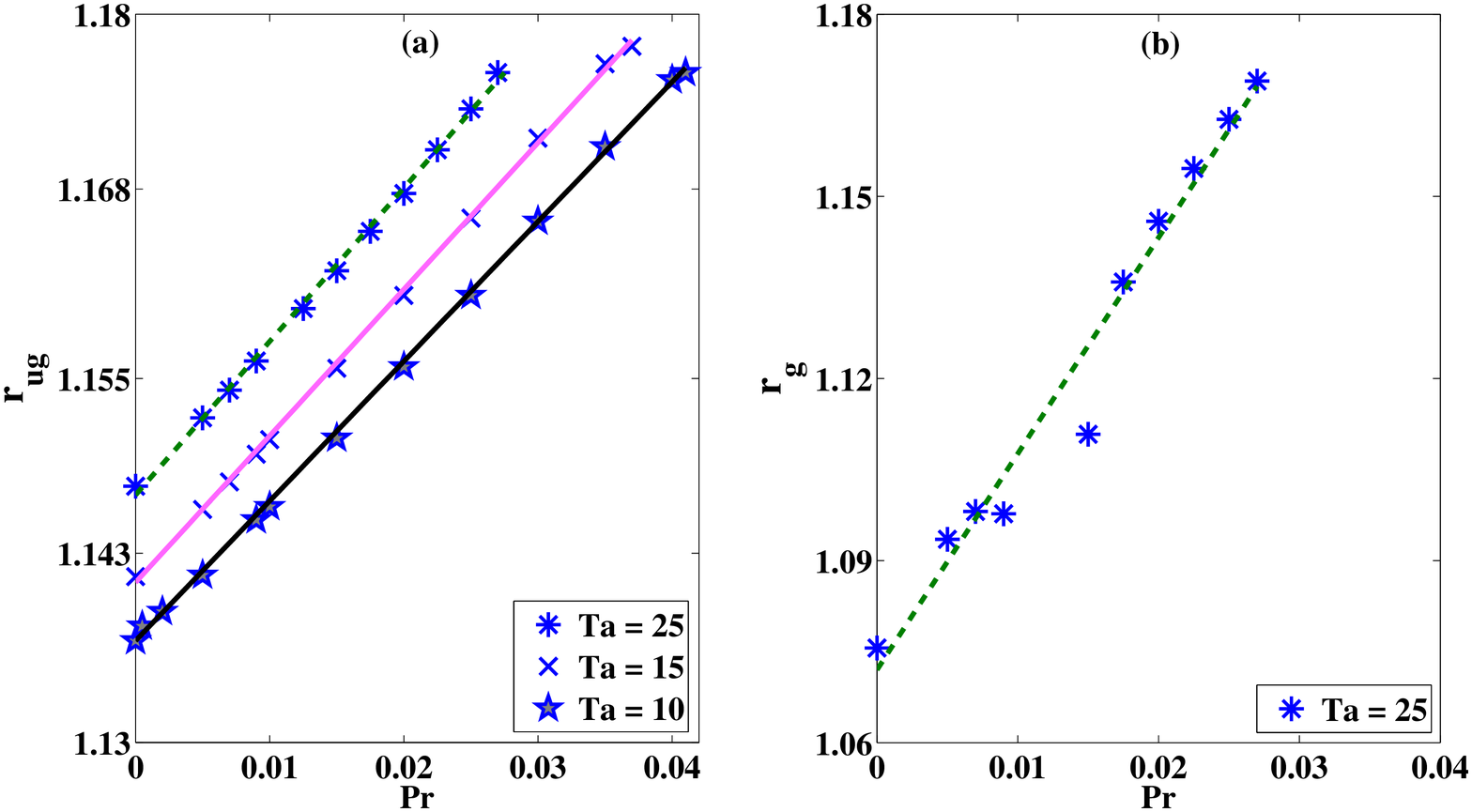}
\caption{Variation of homoclinic bifurcation points with Pr for a fixed value of Taylor number as obtained from model: The variation of homoclinic ungluing point $r_{ug}$ with Prandtl number for three different values of Taylor number (Ta = 10, 15, and 25) are plotted in (a). The points marked by `$\star$', `$\times$', and `$\ast$' denotes the data for Ta = 10, Ta = 15, and Ta = 25 respectiely. The variation of homoclinic gluing bifurcation point $r_g$ with Pr for Ta = 25 is plotted in (b). }
\label{fig:r_Pr}
\end{figure}
 
At higher value of Ta, we observe presence of both gluing and ungluing bifurcation as seen in DNS. The bifurcation diagram for Pr = 0.025 at Ta = 25 is plotted in fig.~\ref{fig:bif_model}(b). WE presence of OCR-II$^{\prime}$ state in addition to all the other states observed for Ta = 10 case. The model shows chaos at the onset while STO solutions were observed in DNS. The first ordered state in the model appears as the OCR-II$^{\prime}$($|W_{101}| = |W_{011}|$) solutions. The two branches of  OCR-II$^{\prime}$ state are marked by the light blue (light gray) dashed-dotted lines. The SQ fixed point solution to be stable for $1.205 \leq r \leq 1.25$ and becomes saddle fixed points $1.205 < r$. Whereas the CR solutions are stable for $1.184 \leq r \leq 1.204$ and becomes unstable for $1.184 < r$. The two branches of OCR-II$^{\prime}$ comes nearer to the saddle fixed point as r is increases in small steps. The one of the branches touch the saddle point the two limit cycle of OCR-II$^{\prime}$ state simultaneously becomes homoclinic orbit to the saddle point. The two limit cycle then spontaneously merge to form a bigger limit cycle which now avoids the saddle point and the system reaches OCR-I state. The transition from OCR-II$^{\prime} \rightarrow$ OCR-I state occurs via homoclinic gluing bifurcation when r is raised above the critical value $r_g$ (represented as the blue (black) square in fig.~\ref{fig:bif_model}). All the other bifurcations are similar as in case of Ta = 10. Variation of homoclinic bifurcation points $r_g$ and $r_{ug}$ with Prandtl number is plotted in fig.~\ref{fig:r_Pr}. Both the homoclinic bifurcation points ($r_{ug}$ and $r_g$) show linear dependence on Pr. The point $r_{ug}$ increases linearly with Pr with a slope close to 1 (fig.~\ref{fig:r_Pr} (a)), while $r_{g}$ varies with a slope equal to 3.65 (fig.~\ref{fig:r_Pr} (b)). 

\section{Conclusions}
We have investigated the effect of Coriolis force on homoclinic bifurcation in RBC. We have found existence of ungluing bifurcation for all values of Taylor numbers, while gluing bifurcation exists on for higher valued of Ta. The values of homoclinic bifurcations points increases as Taylor number is increased at a fixed value of Prandtl number. No gluing or ungluing bifurcation is observe in DNS for Prandtl number greater than 0.04. We have also derived a low dimensional model which can capture the homoclinic bifurcations quite well and n addition can also explain the bifurcations occurring in DNS for higher values of r. The model shows that the homoclinic bifurcations points have linear dependence on the Prandtl number for a fixed values of Ta. We do not observe any homoclinic bifurcations in the model for $0.041 \leq Pr$.

\acknowledgements
We benefited from fruitful discussions with Pinaki Pal, H. Pharasi and A. Basak.

\end{document}